\documentclass[aps,prd, nofootinbib, notitlepage,onecolumn]{revtex4-1}
\usepackage{bm}
\usepackage{latexsym}
\usepackage{amsmath,amsfonts,amssymb}
\usepackage{graphicx,epsfig}
\usepackage{psfrag}
\usepackage{amsthm}
\usepackage{color}
\usepackage{braket}
\usepackage{indent first}
\usepackage[normalem]{ulem}
\usepackage{fancyhdr}
\usepackage{hyperref}
\usepackage[titletoc,title]{appendix}
 \usepackage{environ}
\usepackage{mathtools}
\usepackage{float}
\usepackage[makeroom]{cancel}
\newcommand{\be}{\begin{equation}}
\newcommand{\ee}{\end{equation}}
\newcommand{\bea}{\begin{eqnarray}}
\newcommand{\eea}{\end{eqnarray}}

\NewEnviron{eqn}{
\begin{align}
\begin{split}
  \BODY
\end{split}
\end{align}
}

\NewEnviron{eqn*}{
\begin{align*}
\begin{split}
  \BODY
\end{split}
\end{align*}
}

\begin{document}
\title{Baryon Asymmetry from the Generalized Uncertainty Principle}

\author{Saurya Das} 
\email{saurya.das@uleth.ca}
\affiliation{Theoretical Physics Group and Quantum Alberta, Department of Physics and Astronomy,
University of Lethbridge,
4401 University Drive, Lethbridge,
Alberta, T1K 3M4, Canada}

\author{Mitja Fridman}
\email{fridmanm@uleth.ca}
\affiliation{Theoretical Physics Group and Quantum Alberta, Department of Physics and Astronomy,
University of Lethbridge,
4401 University Drive, Lethbridge,
Alberta, T1K 3M4, Canada}

\author{Gaetano Lambiase} \email{lambiase@sa.infn.it}
\affiliation{Dipartimento di Fisica E.R: Caianiello, Universita di Salerno, Via Giovanni Paolo II, 132 - 84084 Fisciano, Salerno, Italy \& INFN - Gruppo Collegato di Salerno, Italy}

\author{Elias C. Vagenas} \email{elias.vagenas@ku.edu.kw}
\affiliation{Theoretical Physics Group, Department of Physics, Kuwait University, P.O. Box 5969, Safat 13060, Kuwait}

\begin{abstract}
\par\noindent
The unexplained observed baryon asymmetry in the Universe is a long-standing problem in physics, with no satisfactory resolution so far. To explain this asymmetry, three Sakharov conditions must be met.
An interaction term which couples space-time and the baryon current is considered, which satisfies the first two Sakharov conditions.
Furthermore, it is shown that the Generalized Uncertainty Principle (GUP) from quantum gravity induces corrections to the
Friedmann equations in cosmology, via the holographic principle. GUP also induces variations of energy and pressure density in the radiation dominated era, which satisfies the third Sakharov condition. Therefore, this construction 
provides a viable explanation for the observed baryon asymmetry. This also fixes the GUP parameters to $\alpha_0\approx10^4$ and $\beta_0\approx-10^8$.

%
\end{abstract}
\maketitle
%
\section{Introduction}
%
%
%
%
\par\noindent
The two main theories which describe Nature on the smallest and largest scales are Quantum Theory and the Theory of General Relativity, respectively. They are very successful in predicting observable phenomena to a high degree of precision at small and large length scales, respectively \cite{GF,CMW}. 
However, near the Planck energy scale, one expects both quantum and gravitational effects to come into play, requiring a theory of Quantum Gravity (QG) to describe them. 
%
%
%
\par\noindent
While there are a number of promising theories of QG, none has been tested so far, due to limitations of current experiments, and the impossibility of reaching the Planck energy scale with existing technologies. However, Planck scale effects are believed to be universal, which means they should also be present at low and accessible energies,  although their magnitudes are expected to be small \cite{Das:2008kaa,Das:2010sj}. 
Therefore, if precise enough experiments are designed, such effects should be observable. In light of this, Quantum Gravity Phenomenology has become an important area of study, where a lot of phenomenological work has been done so far \cite{CM1,CM2,CM3,CM4,CM5,Buoninfante:2019fwr,Blasone:2019wad,CM6,CM7}.
\par\noindent
One of the main conclusions from QG theories and the pertinent thought experiments is the existence of a minimal measurable length. This means that space-time is believed to be fundamentally discrete. One of the phenomenological models, which describes such discreteness is the Generalized Uncertainty Principle (GUP), which has been studied in great detail \cite{GUP1,GUP2,GUP3,GUP4,KMM,FC,Das:2009hs, Ali:2010yn,ADV0,IPK,Scardigli:2014qka,SLV,KP,VenezGrossMende,Basilakos:2010vs, Das:2020ujn,CM8GS,Iorio:2019wtn,luciano}.
The most general form of 
GUP in three spatial dimensions with linear plus quadratic terms in momentum is given by the commutator \cite{ADV0}
\begin{equation}
    \label{gup}
     [x_i,p_j]=i\hbar\left(\delta_{ij}-\alpha\left(p\delta_{ij}+\frac{p_ip_j}{p}\right)+\beta\left(p^2\delta_{ij}+3p_ip_j\right)\right)
\end{equation}
where $\alpha \equiv \alpha_0/(M_{P}c)$, 
$\beta \equiv \beta_0/(M_{P}c)^2$, 
$\alpha_0$, $\beta_0$ are the dimensionless
linear and quadratic GUP parameters and $M_{P}=\sqrt{\hbar c/G}$ 
is the Planck mass, comprised of the reduced Planck constant $\hbar$, the speed of light in vacuum $c$ and the gravitational constant $G$. With the inclusion of the dimensionless parameters $\alpha_0$ and $\beta_0$, Eq.(\ref{gup}) describes any length scales $\alpha_0\, \ell_P$ and $\sqrt{\beta_0}\,\ell_P$ between the electroweak scale, i.e., $\ell_{EW}\approx10^{-18}\,\mathrm{m}$,  and the Planck scale, i.e., $\ell_P=\sqrt{\hbar G/c^3}\approx10^{-35}\,\mathrm{m}$. For convenience, from here on, natural units, i.e., $\hbar=k_B=c=1$, are used.
\par\noindent
The origin of the baryon asymmetry in the early Universe is an unsolved problem up to this day. Observational evidence shows that the Universe is mostly made up of matter, rather than equal amounts of matter and anti-matter, as is expected by the Quantum and Relativistic theories \cite{CDS}. For such asymmetry to occur, three necessary conditions, called Sakharov conditions \cite{Sakh}, must be met: 1) Baryon number violation, 2) C and CP violation, 3) Deviation from thermal equilibrium.
It should be pointed out that the Cosmic Microwave Background (CMB) temperature anisotropies provide a strong probe of the baryon asymmetry since the observation of the acoustic peaks in CMB and the measurements of large scale structures allow to infer an estimation of the baryon asymmetry parameter $\eta$ (defined as the difference in baryon and anti-baryon densities per unit entropy density; see Eq.(\ref{asym1}) for details) given by $\eta^{(CMB)}\sim (6.3\pm 0.3)\times 10^{-10}$ \cite{27GL}.
On the other hand, measurement of $\eta$ can be also carried out in the context of the Big Bang Nucleosynthes (BBN). The estimation of $\eta$ in such a case gives $\eta^{(BBN)}\sim (3.4-6.9)\times 10^{-10}$ \cite{28GL},
a value that is compatible with the CMB measurement, although derived in two different eras of the Universe evolution.

Although several explanations for the observed baryon asymmetry have been offered so far \cite{CDS,CKB,APS},  none of them has been confirmed yet.
This work explores the possibility that the baryon asymmetry could arise  due to the presence of a minimal measurable length, described by GUP.
It has been shown in Refs. \cite{Cai:2008ys,Zhu:2008cg,AA,Giardino:2020myz} that GUP can modify the Friedmann equations, through modifying the Bekenstein-Hawking entropy, using the holographic principle.  
This approach is used in this work to derive the exact modified Friedmann equations using linear plus quadratic GUP. This in turn provides a general mechanism that allows one to make QG predictions at cosmological scales. An interaction term which couples space-time and baryon current is used to satisfy the first two Sakharov conditions, while the GUP-modified Friedmann equations break thermal equilibrium to satisfy the third Sakharov condition. In contrast to other phenomenological studies, this prediction already has an observational counterpart, and thus offers an explanation for a measured and established feature of Nature, i.e., the baryon asymmetry in the Universe. In addition, this work fixes the values for the dimensionless GUP parameters to $\alpha_0\approx10^4$ and $|\beta_0|\approx10^8$, and thus determines a possible QG length scale of $\ell_{QG}\approx10^{-31}\,\mathrm{m}$. 

This paper is structured as follows. In section \ref{se2}, a modification of the Bekenstein-Hawking entropy formula, due to GUP, is presented. In section \ref{se3}, the modification of the Friedmann equations, using the modified Bekenstein-Hawking entropy, is described. The baryon asymmetry is then explained in section \ref{se4}, by using the modifications derived in previous chapters. In section \ref{se5}, a summary with concluding remarks is given.
%
%
%
%
%
\section{Modification of the Bekenstein-Hawking entropy}
\label{se2}
%
%
%
%
%
On the path to explaining the origin of the baryon asymmetry in the early Universe, one must first modify the Friedmann equations, since such modifications lead to energy density and pressure variations. These variations
break thermal equilibrium and produce the asymmetry. In order to modify the Friedmann equations, one uses the holographic principle, where the Bekenstein-Hawking thermodynamics is used to derive them. The holographic principle states that a description of a theory inside a $d$-dimensional volume in space can be encoded in its $(d-1)$-dimensional boundary, such as an event horizon of a black hole, or the cosmic horizon \cite{GTH,LS}. The holographic principle provides a mechanism to include quantum corrections to large scale systems.

The modification of the Friedmann equations was made in Ref. \cite{AA}, using only the quadratic GUP, i.e.,  the GUP which depends only on the term quadratic  in momentum. 
To examine the conclusions of a
more general model, in this work, we include a GUP which has terms both linear and quadratic in momentum. 
The corresponding commutator is given by Eq.(\ref{gup}). 
The corresponding modified uncertainty relation, derived using
$\Delta x\Delta p\geq\frac{1}{2}|\langle[x,p]\rangle|$ is given by \cite{GUPAM}
%
%
%
%
%
\begin{eqnarray}
    \Delta x\Delta p&\gtrsim& \left[1-2\alpha\langle p\rangle+4\beta\langle p^2\rangle\right] \nonumber \\
    &\geq&\left[1+\left(\frac{\alpha}{\sqrt{\langle p^2\rangle}}+4\beta\right)\Delta p^2+4\beta\langle p\rangle^2-2\alpha\sqrt{\langle p^2\rangle}\right]~,
\end{eqnarray}
%
%
%
\par\noindent
where the prefactor has been set to be of the order $\mathcal{O}(1)$, as seen in Refs. \cite{MV,ACAP,LJG}. Considering the momentum uncertainty $\Delta p=\sqrt{\langle p^2\rangle-\langle p\rangle^2}$, one assumes $\langle p\rangle=0$, as argued in Ref. \cite{KMM}, to obtain the maximal momentum uncertainty and therefore the absolutely smallest possible uncertainty in position. This assumption also implies $\langle p^2\rangle=\Delta p^2$. The modified uncertainty principle then reads
\begin{equation}
\label{upg}
    \Delta x\Delta p\gtrsim\left[1-\alpha\Delta p+4\beta\Delta p^2\right]~.
\end{equation}

%
%
%
%
\par\noindent
As a particle gets absorbed by an apparent horizon, it will change the area of that horizon while the total energy inside the horizon increases \cite{GUP4,Adler:2001vs,MV}. A particle, whose uncertainty is governed by the GUP as given by Eq.(\ref{upg}), will produce a GUP-corrected change to the horizon area. A particle obeying the above uncertainty will have energy proportional to $\Delta p$ \cite{GUP4,Adler:2001vs}.
From Eq.(\ref{upg}), one obtains
%
%
%
%
%
\begin{eqnarray}
\label{delp}
    \Delta p\gtrsim\frac{\Delta x+\alpha}{8\beta}\left(1-\sqrt{1-\frac{16\beta}{\Delta x^2+2\alpha\Delta x+\alpha^2}}\right),
\end{eqnarray}
%
%
%
%
\par\noindent
where the negative solution has been chosen, because one wants to obtain the smallest change in area of the apparent horizon and also this solution is the only one which reduces to the standard Heisenberg uncertainty for $\alpha,\beta\longrightarrow0$. Therefore, the energy of a particle, which gets absorbed by the apparent horizon is given as $E=\Delta p$, where $\Delta p$ is given in Eq.(\ref{delp}).
%
%
%
%
%
%
%
%
\par\noindent
The area of an apparent horizon, which absorbs a particle with energy $E$ and size $R$ changes by $\Delta A\geq8\pi\ell_P^2ER$ \cite{DC,CR} and the particle itself implies the existence of a finite size by $R=\Delta x$, which gives the minimal change of the area of the apparent horizon as
\begin{eqnarray}
\Delta A_{min}\geq8\pi\ell_P^2E\,\Delta x~.
\end{eqnarray}
\par\noindent
The minimal change of the area of an apparent horizon, considering $\Delta x=2r_S$ 
($r_S=2GM$ is the Schwarzschild radius) and $\Delta x^2=A/\pi$ ($A$ is the area of the apparent horizon), is then given as
\begin{eqnarray}
    \Delta A_{min}\simeq\lambda\frac{\ell_P^2(A+\alpha\sqrt{\pi}A^{1/2})}{\beta}\left(1-\sqrt{1-\frac{16\pi\beta}{A+2\alpha\sqrt{\pi}A^{1/2}+\alpha^2\pi}}\right),
\end{eqnarray}
%
%
%
%
\par\noindent
where $\lambda$ is a parameter determined by the Bekenstein-Hawking entropy formula 
$b/\lambda=2\pi$, where 
$b=\Delta S_{min}=\ln{2}$ is the minimal increase in entropy, corresponding to one bit of information \cite{AA}. Therefore, the minimal change of entropy for a minimal change in an apparent horizon area  reads
\begin{eqnarray}
\label{dventropy}
    \frac{\mathrm{d}S}{\mathrm{d}A}=\frac{\Delta S_{min}}{\Delta A_{min}}=\frac{\beta^*}{8\ell_P^2\left(A+\alpha^*A^{1/2}-\sqrt{A^2+2\alpha^*A^{3/2}+(\alpha^{*2}-\beta^*)A}\right)}~,
\end{eqnarray}
%
%
%
\par\noindent
where  
$\alpha^*=\sqrt{\pi}\alpha$ and $\beta^*=16\pi\beta$.
The standard result for the entropy of a black hole (and for any region of space with a horizon, as proposed by the holographic principle) is \cite{JB,SH}
\begin{eqnarray}
\label{bhe}
S=\frac{A}{4G}=\frac{A}{4\ell_P^2}~.
\end{eqnarray}
%
%
%
\par\noindent
 In general, if one wants to modify the entropy, the area $A$ will become a function of $A$, i.e., $f(A)$, \cite{Cai:2008ys}. This entropy can then be written as 
 $S=\tfrac{f(A)}{4\ell_P^2}$. By calculating the derivative of this entropy over area $A$, one obtains
\begin{eqnarray}
\label{gdventropy}
    \frac{\mathrm{d}S}{\mathrm{d}A}=\frac{f'(A)}{4\ell_P^2}~,
\end{eqnarray}
%
%
%
%
\par\noindent
where $f'(A)$ can be identified by comparing the above derivative with  the one in Eq.(\ref{dventropy}), and thus
\begin{eqnarray}
\label{dfa}
    f'(A)=\frac{1}{2}\frac{\beta^*}{\left(A+\alpha^*A^{1/2}-\sqrt{A^2+2\alpha^*A^{3/2}+(\alpha^{*2}-\beta^*)A}\right)}~.
\end{eqnarray}
It should be pointed out that one obtains the standard result $f'(A)=1$ for $\alpha^*,\beta^*\longrightarrow0$. The GUP-modified Bekenstein-Hawking entropy is obtained by integrating Eq.(\ref{dventropy}) over $A$ to obtain
\begin{eqnarray}
S &=& \frac{1}{8\ell_P^2}\left[A\left(1+\sqrt{1+2\alpha^*\frac{1}{A^{1/2}}+(\alpha^{*2}-\beta^*)\frac{1}{A}}\right)+\alpha^*A^{1/2}\left(2+\sqrt{1+2\alpha^*\frac{1}{A^{1/2}}+(\alpha^{*2}-\beta^*)\frac{1}{A}}\right)\right. \nonumber \\
&-&\!\!\!\left.\beta^*\ln{\left(1+\frac{A^{1/2}}{\alpha^*}\left(1+\sqrt{1+2\alpha^*\frac{1}{A^{1/2}}+(\alpha^{*2}-\beta^*)\frac{1}{A}}\right)\right)}\right]~,
\end{eqnarray}
where the standard result from Eq.(\ref{bhe}) is obtained for $\alpha^*,\beta^*\longrightarrow0$. The above equation provides the linear and quadratic GUP-corrected Bekenstein-Hawking entropy. 
%
%
%
%
%
\section{Modification of Friedmann equations}
\label{se3}
%
%
%
%
%
\par\noindent
The modification of the Bekenstein-Hawking entropy obtained in the previous section is necessary to modify the Friedmann equations through the holographic principle. The modification of Friedmann equations to second order in $\alpha^*$ and first order in $\beta^*$ were considered separately 
(and perturbatively) in 
Ref. \cite{Giardino:2020myz}. In this work on the other hand, the modifications of Friedmann equations are calculated exactly and corrections due to $\alpha^*$ and $\beta^*$ are considered simultaneously. One starts with the standard FLRW metric, incorporating homogeneity and isotropy, in a 
$(n+1)-$dimensional space-time
\begin{eqnarray}
\label{metric}
\mathrm{d}s^2=h_{cd}\mathrm{d}x^c\mathrm{d}x^d+\tilde{r}^2\mathrm{d}\Omega_{n-1}^2~,
\end{eqnarray}
\par\noindent
where $h_{cd}={\text{diag}}(-1,a^2/(1-kr^2)
)$, 
$x^c=(t,r)$, $\tilde{r}=a(t)\,r$, 
$\mathrm{d}\Omega_{n-1}$ is the angular part of the $(n-1)$-dimensional sphere, $a=a(t)$ the scale factor, $r$ the comoving radius and $k$ the spatial curvature constant. Indices $c$ and $d$ can only take values $0$ and $1$.
The Friedmann equations, which govern the dynamics of the scale factor $a(t)$, and hence the evolution of the Universe,
undergo modifications depending on the precise form of the function $f(A)$ introduced in the previous section. 
$f(A)=A$ of course gives the standard Friedmann equations,
and deviations from $f(A)=A$ can arise due to QG effects.
%
%
The derivation of the modified Friedmann equations
for an arbitrary $f(A)$ 
is done by considering the holographic principle and the first law of thermodynamics. A detailed derivation is presented in Appendix \ref{feder}. In a space-time with a metric in Eq.(\ref{metric}), the modified Friedmann equations read
%
%
%
%
\begin{eqnarray}
\label{fe1}
    -\frac{8\pi G}{n-1}\left(\rho+p\right)=\left(\dot{H}-\frac{k}{a^2}\right)f'(A)~,
\end{eqnarray}
%
%
%
\par\noindent
and
%
%
%
%
\begin{eqnarray}
\label{fe2}
    -\frac{8\pi G}{n(n-1)}\rho=\frac{\left(n\Omega_n\right)^{\frac{n+1}{n-1}}}{n(n-1)\Omega_n}\int f'(A)\frac{\mathrm{d}A}{A^{\frac{n+1}{n-1}}}~.
\end{eqnarray}
\par\noindent
The above two Friedmann equations are written in the most general form, given an arbitrary functional form $f(A)$. 
At this point, it should be stressed that to obtain the standard case, one has to set $f(A)=A$.  The above modifications are  quantum in nature, since they come from the Bekenstein-Hawking entropy, which is of quantum origin. 
To obtain the GUP-corrected Friedmann equations, one needs to plug Eq.(\ref{dfa}) in Eqs.(\ref{fe1}) and  (\ref{fe2}).
This yields, for the linear + quadratic GUP, the following most general and exact form of the modified Friedmann equations in $n=3$ spatial dimensions, expressed in terms of the standard cosmological parameters $H$, $a$, and $k$ as
%
%
%
%
%
%
%
%
%
%
%
%
\begin{eqnarray}
\label{emfe1}
-4\pi G\left(\rho+p\right)&=&\left(\dot{H}-\frac{k}{a^2}\right) \nonumber \\
&\times& \frac{\beta^*}{8\pi}\frac{\left(H^2+\frac{k}{a^2}\right)}{1+\frac{\alpha^*}{(4\pi)^{1/2}}\left(H^2+\frac{k}{a^2}\right)^{1/2}-\sqrt{1+\frac{2\alpha^*}{(4\pi)^{1/2}}\left(H^2+\frac{k}{a^2}\right)^{1/2}+\frac{(\alpha^{*2}-\beta^*)}{4\pi}\left(H^2+\frac{k}{a^2}\right)}}~ 
\end{eqnarray}
%
%
%
and
\begin{eqnarray}
\label{emfe2}
\frac{8\pi G}{3}(\rho-\Lambda ) &=& \frac{1}{2}\left(H^2+\frac{k}{a^2}\right)+\frac{\alpha^*}{3(4\pi)^{1/2}}\left(H^2+\frac{k}{a^2}\right)^{3/2}+\frac{2\pi (\alpha^{*2}+2\beta^*)}{3(\alpha^{*2}-\beta^*)^2}\nonumber \\
&+&\left[\frac{1}{3}\left(H^2+\frac{k}{a^2}\right)+\frac{(4\pi)^{1/2} \alpha^*}{6(\alpha^{*2}-\beta^*)}\left(H^2+\frac{k}{a^2}\right)^{1/2}-\frac{2\pi(\alpha^{*2}+2\beta^*)}{3(\alpha^{*2}-\beta^*)^2}\right] \nonumber \\
 &\times& \sqrt{1+\frac{2\alpha^*}{(4\pi)^{1/2}}\left(H^2+\frac{k}{a^2}\right)^{1/2}+\frac{(\alpha^{*2}-\beta^*)}{4\pi}\left(H^2+\frac{k}{a^2}\right)} \nonumber \\
&+& \frac{2\pi\alpha^*\beta^*}{(\alpha^{*2}-\beta^*)^{5/2}}\ln{\left[1+\frac{(\alpha^{*2}-\beta^*)}{(4\pi)^{1/2}(\alpha^*+\sqrt{\alpha^{*2}-\beta^*})}\left(H^2+\frac{k}{a^2}\right)^{1/2}\right.} \nonumber \\
&+&{\left.\frac{\sqrt{\alpha^{*2}-\beta^*}}{\alpha^*+\sqrt{\alpha^{*2}-\beta^*}}\left(\sqrt{1+\frac{2\alpha^*}{(4\pi)^{1/2}}\left(H^2+\frac{k}{a^2}\right)^{1/2}+\frac{(\alpha^{*2}-\beta^*)}{4\pi}\left(H^2+\frac{k}{a^2}\right)}-1\right)\right]}~, 
\end{eqnarray}
%
%
%
%
%
%
where
\begin{eqnarray}
A=4\pi \tilde{r}_A^2=\frac{4\pi}{H^2+\frac{k}{a^2}}~
\end{eqnarray}
was used. In the above $\tilde{r}_A$ is the radius of the apparent horizon and is defined in Eq.(\ref{aprad}) in Appendix \ref{feder}. For vanishing GUP parameters $\alpha^*,\beta^*\longrightarrow0$ in Eqs. (\ref{emfe1}) and (\ref{emfe2}), one obtains the standard Friedmann equations. Details of this limit are given in Appendix \ref{limit}.
%
%
%
%

It is evident that if one wants to study the radiation-dominated era, the tiny observed cosmological constant $\Lambda$ will be ignored, and the spatial curvature constant has to be set $k=0$, consistent with the observed spatially flat Universe. As a result the GUP-modified Friedmann equations in Eqs.(\ref{emfe1}) and (\ref{emfe2}) are further simplified, respectively,  to
%
%
%
%
\begin{eqnarray}
\label{cmfe1}
-4\pi G\left(\rho+p\right)=\frac{\beta^*\dot{H}}{8\pi}\frac{H^2}{1+\frac{\alpha^*}{(4\pi)^{1/2}}H-\sqrt{1+\frac{2\alpha^*}{(4\pi)^{1/2}}H+\frac{(\alpha^{*2}-\beta^*)}{4\pi}H^2}}~
\end{eqnarray}
%
%
%
%
and
\begin{eqnarray}
\label{cmfe2}
\frac{8\pi G}{3}\rho&=&\frac{1}{2}H^2+\frac{\alpha^*}{3(4\pi)^{1/2}}H^3+\frac{2\pi(\alpha^{*2}+2\beta^*)}{3(\alpha^{*2}-\beta^*)^2}\nonumber \\
&+&\left[\frac{1}{3}H^2+\frac{(4\pi)^{1/2}\alpha^*}{6(\alpha^{*2}-\beta^*)}H-\frac{2\pi(\alpha^{*2}+2\beta^*)}{3(\alpha^{*2}-\beta^*)^2}\right]\sqrt{1+\frac{2\alpha^*}{(4\pi)^{1/2}}H+\frac{(\alpha^{*2}-\beta^*)}{4\pi}H^2} \nonumber \\
&+&\frac{2\pi\alpha^*\beta^*}{(\alpha^{*2}-\beta^*)^{5/2}}\ln{\left[1+\frac{(\alpha^{*2}-\beta^*)}{(4\pi)^{1/2}(\alpha^*+\sqrt{\alpha^{*2}-\beta^*})}H\right.} \nonumber \\
&+&{\left.\frac{\sqrt{\alpha^{*2}-\beta^*}}{\alpha^*+\sqrt{\alpha^{*2}-\beta^*}}\left(\sqrt{1+\frac{2\alpha^*}{(4\pi)^{1/2}}H+\frac{(\alpha^{*2}-\beta^*)}{4\pi}H^2}-1\right)\right]}~.
\end{eqnarray}
%
%
%
%
\par\noindent
These GUP-modified Friedmann equations form the basis on which energy density and pressure variations in the early universe are studied. 
In what follows, they will also be used to break 
thermodynamic equilibrium and explain the baryon asymmetry in the early universe.
%
%
%
%
\section{Gravitational Baryon asymmetry}
\label{se4}
%
%
%
%
%
\par\noindent
All three Sakharov conditions, listed in the introduction, must be met to explain the observed baryon asymmetry. As seen from the following considerations, the first two Sakharov conditions are satisfied by introducing a coupling term, which couples space-time to the baryon current and the final Sakharov condition is satisfied by breaking thermal equilibrium through the modified Friedmann equations, derived in the previous section.

Within supergravity theories, a mechanism for generating baryon asymmetry during the expansion of the universe by means of a dynamical breaking of CPT (and CP) has been studied \cite{KU1,KU2}. However, in this mechanism, the thermal equilibrium is preserved, thus not all the Sakharov conditions are satisfied. The interaction responsible for the  CPT violation is given by a coupling between the derivative of the Ricci scalar curvature $R$ and the baryon current $J^\mu$ \cite{gravbar} 
\begin{eqnarray}
\label{intterm}
\frac{1}{M_*^{2}}\int\mathrm{d}^4x\sqrt{-g}\,J^\mu\partial_\mu R\,,
\end{eqnarray}
where $M_*$ is the cutoff scale characterizing the effective theory (see Ref. \cite{allGB} for further applications). 
If there exist interactions that violate the baryon number $B$ in thermal equilibrium, to satisfy the first Sakharov condition, then a net baryon asymmetry can be generated and gets frozen-in below the decoupling temperature\footnote{During the evolution of the Universe, the CPT violation generates the baryon asymmetry (B-asymmetry). This occurs when baryon (or lepton) violating interactions are still in thermal equilibrium. The asymmetry is frozen at the decoupling temperature $T_D$, when the baryon (or lepton) violation goes out of equilibrium. The temperature $T_D$ is derived from the relation $\Gamma(T_D) \simeq H(T_D)$, where $\Gamma$ is the interaction rate of processes and $H$ the expansion rate of the Universe. More specifically, in the regime $\Gamma \gg H$, or $T> T_D$, the B-asymmetry is generated by B-violating processes at thermal equilibrium; at $T=T_D$, i.e., $\Gamma \simeq H$, the decoupling occurs, while when $\Gamma < H$, or $T < T_D$ the B-asymmetry gets frozen.}  %
$T_D$. By taking the integrand from Eq.(\ref{intterm}), i.e., the Lagrangian density, and noting that the spatial part of $\partial_\mu R$ vanishes for the FLRW metric, one gets
\begin{eqnarray}
\label{integrand}
\frac{1}{M_*^2}J^\mu\partial_\mu R=\frac{1}{M_*^2}(n_B-n_{\Bar{B}})\dot{R}~.
\end{eqnarray}
\par\noindent
The effective chemical potential for baryons, i.e., $\mu_B$, and for anti-baryons, i.e., $\mu_{\Bar{B}}$, can be read from the equation above as
\begin{eqnarray}
\label{chempot}
\mu_B=-\mu_{\Bar{B}}=-\frac{\dot{R}}{M_*^2}~,
\end{eqnarray}
%
%
%
%
since Eq. (\ref{integrand}) corresponds to the energy density term for a grand canonical ensemble. For relativistic particles, the net  baryon number density of matter in the early Universe is given by \cite{KT} 
\begin{eqnarray}
n_B-n_{\Bar{B}}=\frac{g_b}{6}\mu_BT^2~,
\end{eqnarray}
\par\noindent
where $g_b\sim\mathcal{O}(1)$ is the number of intrinsic degrees of freedom of baryons. The baryon asymmetry is defined in the standard notation as \cite{KT}
\begin{eqnarray}
\label{base}
\frac{1}{7}\eta\equiv\frac{n_B-n_{\Bar{B}}}{s}\approx\frac{n_B}{s}\simeq-\frac{15\,g_b}{4\pi^2g_*}\frac{\dot{R}}{M_*^2T}\bigg|_{T_D}~,
\label{asym1}
\end{eqnarray}
%
%
%
where $s=\frac{2\pi^2g_{*s}}{45}T^3$ is the entropy per unit volume, i.e., entropy density, in the radiation-dominated era, and $g_{*s}$ is the number of degrees of freedom for particles which contribute to the entropy of the Universe. It may be noted that $g_{*s}$ takes values very close to the total number of degrees of freedom of relativistic Standard Model particles $g_*$, i.e., $g_{*s} \approx g_*\sim106$, as discussed in Ref. \cite{KT}. 
\par\noindent
The parameter $\eta$ is different from zero provided that the time derivative of the Ricci scalar $\dot{R}$ is nonvanishing. The baryon asymmetry equals to zero ($\eta=0$) without QG corrections, because in that case $\dot{R}=0$ in the radiation-dominated era (in the standard cosmological model), due to thermal equilibrium still being satisfied, as is evident from the following calculation.
\par\noindent
The deviation from thermal equilibrium is described in terms of the variations in energy density and pressure. The energy density and pressure including these variations are written as
\begin{eqnarray}
\label{perturb}
\rho=\rho_0+\delta\rho\,,\qquad p=p_0+\delta p~,
\end{eqnarray}
\par\noindent
 where $\rho_0$ and  $p_0$ are the equilibrium energy density and pressure,  respectively. 
 One substitutes the expressions from Eq.(\ref{perturb}) in the GUP-modified Friedmann equations, namely Eqs.(\ref{cmfe1}) and (\ref{cmfe2}), to obtain the GUP-induced variations as
\begin{eqnarray}
\label{drho}
\delta\rho=\frac{\alpha^*}{3}\sqrt{\frac{8G}{3}}\rho_0^{3/2}-\frac{\beta^*}{12}G\rho_0^2
\end{eqnarray}
%
%
%
and
\begin{eqnarray}
\label{dp}
\delta p=\frac{\alpha^*}{6}\sqrt{\frac{8G}{3}}(1+3w)\rho_0^{3/2}-\frac{\beta^*}{12}G(1+2w)\rho_0^2~,
\end{eqnarray}
%
%
%
%
\par\noindent
where $w$ is defined through the equation of state 
$p_0=w\rho_0$. To obtain the variations in Eqs.(\ref{drho}) and (\ref{dp}), the GUP-modified Friedmann Eqs.(\ref{cmfe1}) and (\ref{cmfe2}) were expanded in a Taylor series up to the fourth order in $\alpha^*$ and second order in $\beta^*$, to obtain all terms up to quadratic order in the GUP parameters. Note here that the variations in Eqs. (\ref{drho}) and (\ref{dp}) are valid for any epoch with $-1\leq w\leq1/3$. For the purpose of this work, $w$ is taken to be constant, since all the relevant physics happens in the radiation-dominated era with $w=1/3$.
\par\noindent
As seen from the following, GUP corrections break thermal equilibrium and modify $\dot{R}$, making it non-vanishing. 
To obtain the GUP-corrected derivative of the Ricci scalar $\dot{R}$, the trace of the Einstein equation is computed as
\begin{eqnarray}
\label{ricci}
R=-8\pi G\,T_g=-8\pi G\,(\rho-3p)~,
\end{eqnarray}
%
%
%
%
%
\par\noindent
where 
$T_g=\rho-3p$ is the trace of the energy-momentum tensor. The GUP-corrected Ricci scalar is obtained by plugging the energy density and pressure from Eq.(\ref{perturb}), with their respective variations from Eqs.(\ref{drho}) and (\ref{dp}), in Eq.(\ref{ricci}) and reads as
\begin{eqnarray}
\label{ricciexp}
R=-8\pi G\,(1-3w)\rho_0+\alpha^*\frac{8\sqrt{2}\pi}{3\sqrt{3}}G^{3/2}(1+9w)\rho_0-\beta^*\frac{4\pi}{3}G^2(1+3w)\rho_0^2~.
\end{eqnarray}
\par\noindent
Next, to compute the time derivative of the Ricci scalar from Eq.(\ref{ricciexp}),
one considers the continuity equation from Eq.(\ref{conteq}) in $n=3$ spatial dimensions
\begin{eqnarray}
\label{conteq3}
\dot{\rho}_0+3H(1+w)\rho_0=0~
\end{eqnarray}
and the equilibrium form of the second Friedmann equation $H^2=\frac{8\pi G}{3}\rho_0$ (as corrections to the latter would only contribute to orders higher than those considered in this paper). 

Even if the constant vacuum energy density $\Lambda$ is not negligible, its time derivative would vanish and the following results would remain unchanged.
Therefore, the GUP-corrected time derivative of the Ricci scalar  turns out to be
\begin{eqnarray}
\label{dtriccigen}
\dot{R}&=&\sqrt{3}(8\pi)^{3/2}G^{3/2}(1-2w-3w^2)\rho_0^{3/2}-\alpha^*16\,\pi^{3/2}G^2(1+w)(1+9w)\rho_0^2 \nonumber \\
&+&\beta^*\frac{(8\pi)^{3/2}}{\sqrt{3}}G^{5/2}(1+3w)(1+w)\rho_0^{5/2}~.
\end{eqnarray}
%
%
%
%
The above derivative of the Ricci scalar is then evaluated at the radiation-dominated era, when $w=1/3$, and reads
\begin{eqnarray}
\label{dtricci}
\dot{R}=-\alpha^*\frac{256}{3}\,\pi^{3/2}\,G^2\rho_0^2+\beta^*\,8\left(\frac{8\pi}{3}\right)^{3/2}\!\!\!G^{5/2}\rho_0^{5/2}~.
\end{eqnarray}
%
%
%
%
From the above one can see, that GUP effects provide an essential mechanism to break the thermal equilibrium, thus satisfying the third and final Sakharov condition. One then substitutes Eq.(\ref{dtricci}) in the baryon asymmetry formula in Eq.(\ref{base}) to obtain
\begin{eqnarray}
\label{bas}
\eta=\alpha_0\frac{112\,\pi^2g_*g_b}{45}\left(\frac{T_D}{M_P}\right)^7\left(\frac{M_P}{M_*}\right)^2-\beta_0\frac{896\sqrt{5}\,\pi^3g_*^{3/2}g_b}{675}\left(\frac{T_D}{M_P}\right)^9\left(\frac{M_P}{M_*}\right)^2~,
\end{eqnarray}
%
%
%
%
where the gravitational constant is expressed in terms of the Planck mass, i.e., $M_P$, as 
$G=1/M_P^2$ and the equilibrium energy density $\rho_0$ is replaced by \cite{KT}
\begin{eqnarray}
\rho_0=\frac{\pi g_*}{30}T^4~.
\end{eqnarray}
%
%
%
%
\par\noindent
In order to obtain an estimate of the GUP parameters, the expression in Eq.(\ref{bas}) has to be evaluated at the decoupling temperature 
$T_D=M_I$  where $M_I\sim 2\times10^{16}\,\mathrm{GeV}$ is the upper bound on the tensor mode fluctuation constraints in the inflationary scale \cite{Kinney:2006qm}, and  the cutoff scale, i.e., $M_*$, is taken to be the reduced Planck mass, namely $M_*=M_P/\sqrt{8\pi}$. Then one obtains
\begin{eqnarray}
\label{basnum}
\eta=\alpha_0\,2.08\times10^{-15}-\beta_0\,2.16\times10^{-19}~.
\end{eqnarray}
%
%
%
%
\par\noindent
Given the measured baryon asymmetry between 
$5.7\times10^{-11}\lesssim\eta\lesssim9.9\times10^{-11}$
\cite{baryogenesis}, the dimensionless GUP parameters can be constrained for four distinct cases. 
\begin{itemize}
    \item Only linear GUP ($\beta_0=0$): $2.74\times10^4\lesssim\alpha_0\lesssim4.76\times10^4$
    \item Only quadratic GUP ($\alpha_0=0$): $-4.58\times10^8\lesssim\beta_0\lesssim-2.64\times10^8$
    \item Linear and Quadratic GUP ($\beta_0=-\alpha_0^2$): $1.21\times10^4\lesssim\alpha_0\lesssim1.71\times10^4$ and $-2.92\times10^8\lesssim\beta_0\lesssim-1.46\times10^8$
    \item Linear and Quadratic GUP: $\alpha_0\gtrsim4.81\times10^3$ and $\beta_0\lesssim-1.48\times10^8$
\end{itemize}
\par\noindent
Based on the existing measured baryon asymmetry, the dimensionless GUP parameters should be between the above values, depending on the model choice. For an arbitrary choice of the model, the parameters $\alpha_0$ and $\sqrt{-\beta_0}$ should be between $1.21\times10^4$ and $4.76\times10^4$, except for the last case, where the lower bounds $\alpha_0\gtrsim4.81\times10^3$ and $\sqrt{-\beta_0}\gtrsim1.22\times10^4$ are obtained. To estimate these bounds $\beta_0=\alpha_0^2+\beta'_0$ was used in Eq. (\ref{basnum}), since $\mathcal{O}(\beta_0)\sim\mathcal{O}(\alpha_0^2)$. Here $\beta'_0$ is a deviation of $\beta_0$ from $\alpha_0^2$. The first three cases set the QG length scale to be $1.21\times10^{-31}\,\mathrm{m}\lesssim\ell_{QG}\lesssim4.76\times10^{-31}\,\mathrm{m}$, depending on the model. It is noteworthy that the dimensionless quadratic GUP parameter $\beta_0$ has a negative value.
%
%
%
%
\section{Conclusion}
\label{se5}
%
%
%
%
\par\noindent
The baryon asymmetry produced in the radiation-dominated era of the Universe can be explained, if there is a mechanism satisfying the three Sakharov conditions. In this paper, it has been shown  that this mechanism  can indeed be achieved within the context of a GUP model under consideration. 
The CP symmetry is broken by the coupling between the derivative of the Ricci scalar and the baryon current, as seen in Eq. (\ref{intterm}). Considering interactions which break the baryon number $B$ (GUT interactions, for example see Ref. \cite{KT}) the thermal equilibrium is broken by GUP. 
Therefore, through this mechanism, all three Sakharov conditions are satisfied.

It was shown that GUP modifies the apparent horizon area, which in turn modifies the Bekenstein-Hawking entropy of the observable Universe. 
Then the first law of thermodynamics and the continuity equation 
provide the GUP-modified Friedmann equations which break thermal equilibrium. From the broken thermal equilibrium one obtains a non-vanishing time derivative of the Ricci scalar, which in turn produces a non-vanishing baryon asymmetry in the early Universe.
It may be noted that the GUP-modified Friedmann  Eqs.(\ref{emfe1}) and (\ref{emfe2}) are exact.
These GUP-modified equations 
can be applied to other cosmological problems as well. 
\par\noindent
Using the above, it was possible to 
obtain strict bounds on the dimensionless GUP parameters, as seen at the end of section \ref{se4}. This sets the QG length scale between $1.21\times10^{-31}\,\mathrm{m}\lesssim\ell_{QG}\lesssim4.76\times10^{-31}\,\mathrm{m}$, with  the exact range to depend on the version of GUP being under study,  e.g. linear vs. quadratic. 
The above range is one of the most stringent that has been obtained so far and it is hoped that experiments and observations in the near future may be able to detect such a scale. It is satisfying to note that the obtained values for the quadratic GUP parameter $\beta_0$ are close to the upper and lower bounds on the parameter found in a recent work by one of the current authors and his collaborators \cite{Nenmeli:2021orl}.
%
%
%
%

Finally it  should be mentioned that a negative value of the GUP parameter, i.e., $\beta<0$, could arise in non-trivial structures of the space-time,  such as the discreteness of space \cite{Buoninfante:2019fwr,JKS,Nenmeli:2021orl,KMM,vilasi,Vagenas:2017fwa,Aghababaei:2021gxe}. A similar result follows also in the context of the crystal lattice \cite{JKS}. This could suggest that, at the level of Planck scale, the space-time could have a lattice or granular structure \cite{Ali:2009zq,Das:2010zf,Das:2020ujn,Deb:2016psq}.

\section{Acknowledgement}

This work was supported by the Natural Sciences and Engineering Research Council of Canada. SD and ECV would like to acknowledge networking support by the COST Action CA18108. We thank G. Luciano for useful discussions and suggestions.

\begin{appendices}

\section{}
\label{feder}

In this appendix a detailed derivation of the first Friedmann equation is given and follows the steps from Ref. \cite{AA}. The second Friedmann equation follows from the first, as explained in the main text in section \ref{se3}.
The main assumptions of the holographic principle are that the entropy of the apparent horizon is that given by Eq.(\ref{bhe})  while the temperature of the apparent horizon  is given by \cite{CK,Cai:2008gw}
\begin{eqnarray}
\label{htemp}
T=\frac{\kappa}{2\pi}~,
\end{eqnarray}
%
where $\kappa=\frac{1}{2\sqrt{-h}}\partial_c\left(\sqrt{-h}h^{cd}\partial_b\tilde{r}\right)=-\frac{1}{\tilde{r}_A}\left(1-\frac{\dot{\tilde{r}}_A}{2H\tilde{r}_A}\right)$ is the surface gravity of the apparent horizon. 
Here $h=\mathrm{det}(h^{cd})$ and $\tilde{r}_A$ is the radius of the apparent horizon. The location of the 
apparent horizon is obtained from the equation 
$h^{cd}\partial_c\tilde{r}\partial_d\tilde{r}=0$ and reads as \cite{CK}
\begin{eqnarray}
\label{aprad}
    \tilde{r}_A=\frac{1}{\sqrt{H^2+\frac{k}{a^2}}}~,
\end{eqnarray}
%
%
%
where $H=\dot{a}/a$ is the Hubble parameter. As usual, one assumes that the matter in the universe is a perfect fluid and is described by the energy-momentum tensor
\begin{eqnarray}
T_{\mu\nu}=\left(\rho+p\right)u_\mu u_\nu+{p}g_{\mu\nu}~,
\end{eqnarray}
%
%
%
\par\noindent
where $u_\mu$ is the four velocity, $\rho$ the energy density, $p$ the pressure and $g_{\mu\nu}$ the space-time metric of the $(n+1)$-dimensional FLRW model. The energy conservation law, i.e.,  $T^{\mu\nu}_{\,\,\,\,\,\,;\nu}=0$, for a perfect fluid gives the continuity equation
\begin{eqnarray}
\label{conteq}
\dot{\rho}+nH\left(\rho+p\right)=0~.
\end{eqnarray}
%
%
%
%
\par\noindent
The Friedmann equations are obtained by considering the first law of thermodynamics for the matter content inside the apparent horizon
\begin{eqnarray}
\label{flt}
\mathrm{d}E=T\mathrm{d}S+W\mathrm{d}V~,
\end{eqnarray}
where $E$ is the total energy inside the apparent horizon, given by $E=\rho V$,  $V$ is the volume of an $n$-dimensional sphere 
which is given by $V=\Omega_n\,\tilde{r}_A^{\,n}$, with $\Omega_n=\tfrac{\pi^{n/2}}{\Gamma(n/2+1)}$ and has an area of $A=n\, \Omega_{n}\,\tilde{r}_{A}^{\,n-1}$, and  $W$ is the work density which reads  as \cite{SAH}
\begin{eqnarray}
\label{wdens}
W=-\frac{1}{2}T^{cd}h_{cd}=\frac{1}{2}\left(\rho-p\right)~.
\end{eqnarray}
%
%
%
\par\noindent
Given all of the above information, now one has to evaluate all the terms in Eq.(\ref{flt}) to obtain the first Friedmann equation. Using Eq.(\ref{conteq}), the energy differential reads
\begin{eqnarray}
\label{dE}
\mathrm{d}E=n\,\Omega_n\tilde{r}_A^{\,n-1}\rho\mathrm{d}\tilde{r}_A-n\,\Omega_n\,\tilde{r}_A^{\,n}(\rho+p)H\mathrm{d}t~,
\end{eqnarray}
where $\mathrm{d}V=n\, \Omega_{n}\,\tilde{r}_{A}^{\,n-1}\mathrm{d}\tilde{r}_A$ has been used. The second term in Eq.(\ref{flt}), using Eqs.(\ref{gdventropy}) and (\ref{htemp}) reads
\begin{eqnarray}
\label{TdS}
T\mathrm{d}S=-\frac{1}{2\pi\tilde{r}_A}\left(1-\frac{\dot{\tilde{r}}_A}{2H\tilde{r}_A}\right)\frac{f'(A)\,n\,(n-1)\,\Omega_n\,\tilde{r}_A^{n-2}}{4G}\,\mathrm{d}\tilde{r}_A~.
\end{eqnarray}
And finally, the third term in Eq.(\ref{flt}), using Eq.(\ref{wdens}) reads 
\begin{eqnarray}
\label{WdV}
W\mathrm{d}V=\frac{1}{2}n\,\Omega_n\,\tilde{r}_A^{n-1}(\rho-p)\,\mathrm{d}\tilde{r}_A~.
\end{eqnarray}
Plugging Eqs. (\ref{dE}), (\ref{TdS}) and (\ref{WdV}) in Eq.(\ref{flt}), one obtains the first Friedmann equation
\begin{eqnarray}
\label{fe1a}
    -\frac{8\pi G}{n-1}\left(\rho+p\right)=\left(\dot{H}-\frac{k}{a^2}\right)f'(A)~.
\end{eqnarray}
%
%
%
\par\noindent
To correctly derive the above Friedmann equation, one must consider $\dot{\tilde{r}}_A=0$, since the apparent horizon radius is assumed to be fixed in an infinitesimal time interval, which constrains the possible equation of state to $p\simeq-\rho$, which also must be used with terms including $(\rho-p)$. Using the continuity equation as given in Eq.(\ref{conteq}) with Eq.(\ref{fe1a}) and integrating it, one obtains the second Friedmann equation
\begin{eqnarray}
\label{fe2a}
    -\frac{8\pi G}{n(n-1)}\rho=\frac{\left(n\Omega_n\right)^{\frac{n+1}{n-1}}}{n(n-1)\Omega_n}\int f'(A)\frac{\mathrm{d}A}{A^{\frac{n+1}{n-1}}}~.
\end{eqnarray}
The Friedmann equations derived in Eqs. (\ref{fe1a}) and (\ref{fe2a}) are presented in the main text in Eqs. (\ref{fe1}) and (\ref{fe2}).

After plugging $f'(A)$ from Eq.(\ref{dfa}) in Eqs. (\ref{fe1a}) and (\ref{fe2a}) for $n=3$ spatial dimensions, one obtains the Friedmann equations, modified by the linear plus quadratic GUP as
\begin{eqnarray}
\label{mfe1}
    -4\pi G\left(\rho+p\right)=\left(\dot{H}-\frac{k}{a^2}\right)\frac{\beta^*}{2}\frac{1}{A+\alpha^*A^{1/2}-\sqrt{A^2+2\alpha^*A^{3/2}+(\alpha^{*2}-\beta^*)A}}
\end{eqnarray}
%
%
%
and
\begin{eqnarray}
\label{mfe2}
    \frac{8\pi G}{3}\rho &=&-4\pi\int \frac{\beta^*}{2A^2}\frac{\mathrm{d}A}{\left(A+\alpha^*A^{1/2}-\sqrt{A^2+2\alpha^*A^{3/2}+(\alpha^{*2}-\beta^*)A}\right)}= 2\pi \Bigg[\frac{1}{A}+\alpha^*\frac{2}{3}\frac{1}{A^{3/2}} \nonumber  \\
     &+&\left(\vphantom{\frac{\alpha^{*2}+2\beta^*}{3(\alpha^{*2}-\beta^*)^2}}\right.\frac{2}{3}\frac{1}{A}+\frac{\alpha^*}{3(\alpha^{*2}-\beta^*)}\frac{1}{A^{1/2}}-\underbrace{{\frac{3\alpha^{*2}-2(\alpha^{*2}-\beta^*)}{3(\alpha^{*2}-\beta^*)^2}}}_{\displaystyle{\frac{\alpha^{*2}+2\beta^*}{3(\alpha^{*2}-\beta^*)^2}}}\left.\vphantom{\frac{\alpha^{*2}+2\beta^*}{3(\alpha^{*2}-\beta^*)^2}}\right)\sqrt{1+2\alpha^{*}\frac{1}{A^{1/2}}+(\alpha^{*2}-\beta^*)\frac{1}{A}} \nonumber \\
    &+& \frac{\alpha^{*}\beta^*}{(\alpha^{*2}-\beta^*)^{5/2}}\ln{\left(\alpha^{*}+(\alpha^{*2}-\beta^*)\frac{1}{A^{1/2}}+\sqrt{\alpha^{*2}-\beta^*}\sqrt{1+2\alpha^{*}\frac{1}{A^{1/2}}+(\alpha^{*2}-\beta^*)\frac{1}{A}}\right)}\Bigg] \nonumber \\
     &+& C
\end{eqnarray}
%
%
%
%
where $C$ is an integration constant. $C$ is determined by considering the boundary conditions in the vacuum energy (dark energy) dominated era, where the energy density goes to $\rho=\rho_{\mathrm{vac}}=\Lambda$ as the area of the apparent horizon of the universe goes to $A\longrightarrow\infty$
\begin{eqnarray}
\label{const}
    C=\frac{8\pi G}{3}\Lambda+2\pi \left(\frac{\alpha^{*2}+2\beta^*}{3(\alpha^{*2}-\beta^*)^2}-\frac{\alpha^{*}\beta^*}{(\alpha^{*2}-\beta^*)^{5/2}}\ln{\left(\alpha^{*}+\sqrt{\alpha^{*2}-\beta^*}\right)}\right)~. 
\end{eqnarray}
%
%

\section{}
\label{limit}

In this Appendix the limit for vanishing GUP parameters $\alpha^*,\beta^*\longrightarrow0$ is studied, to obtain the standard Friedmann equations. For this, we use the Taylor expansions 
$\sqrt{1+x}\approx1+\frac{x}{2}-\frac{x^2}{8}$ and $\ln{(1+x)}\approx x-\frac{x^2}{2}$, i.e., 
up to second order, as required. 
By second order, one means of course that all terms up to those $\propto\alpha^{*2}$ and $\propto\beta^{*}$ are retained and higher order terms are dropped. 
%
%
Here, both terms are considered 
small simultaneously, as they are  proportional to Planck length and its square respectively, 
and therefore in the limit 
$\ell_{P} \longrightarrow 0$, required to obtain standard results, they both approach zero concurrently. 
%
In this limit, the first GUP-modified Friedmann equation from Eq.(\ref{emfe1}) reduces to
\begin{eqnarray}
\label{emfe3}
&-&4\pi G\left(\rho+p\right)\approx\left(\dot{H}-\frac{k}{a^2}\right) \nonumber \\
&\times& \frac{\beta^*}{8\pi}\frac{\left(H^2+\frac{k}{a^2}\right)}{\bcancel{1}+\bcancel{\frac{\alpha^*}{(4\pi)^{1/2}}\left(H^2+\frac{k}{a^2}\right)^{1/2}}-\bcancel{1}-\bcancel{\frac{\alpha^*}{(4\pi)^{1/2}}\left(H^2+\frac{k}{a^2}\right)^{1/2}}-\frac{(\cancel{\alpha^{*2}}-\beta^*)}{8\pi}\left(H^2+\frac{k}{a^2}\right)+\bcancel{\frac{\alpha^{*2}}{8\pi}\left(H^2+\frac{k}{a^2}\right)}+\mathcal{O}(\alpha^{*3})}~ \nonumber \\
&=&\left(\dot{H}-\frac{k}{a^2}\right)\frac{\bcancel{\beta^*}}{\bcancel{8\pi}}\frac{\bcancel{8\pi}}{\bcancel{\beta^*}}\frac{\bcancel{\left(H^2+\frac{k}{a^2}\right)}}{\bcancel{\left(H^2+\frac{k}{a^2}\right)}}=\dot{H}-\frac{k}{a^2}~,
\end{eqnarray}
while the second GUP-modified Friedmann equation from Eq.(\ref{emfe2}) reduces to
\begin{eqnarray}
\label{emfe4}
\frac{8\pi G}{3}(\rho-\Lambda ) &\approx& \frac{1}{2}\left(H^2+\frac{k}{a^2}\right)+\frac{\alpha^*}{3(4\pi)^{1/2}}\left(H^2+\frac{k}{a^2}\right)^{3/2}+\frac{2\pi (\alpha^{*2}+2\beta^*)}{3(\alpha^{*2}-\beta^*)^2}\nonumber \\
&+&\left[\frac{1}{3}\left(H^2+\frac{k}{a^2}\right)+\frac{(4\pi)^{1/2} \alpha^*}{6(\alpha^{*2}-\beta^*)}\left(H^2+\frac{k}{a^2}\right)^{1/2}-\frac{2\pi(\alpha^{*2}+2\beta^*)}{3(\alpha^{*2}-\beta^*)^2}\right] \nonumber \\
 &\times& \left(1+\frac{\alpha^*}{(4\pi)^{1/2}}\left(H^2+\frac{k}{a^2}\right)^{1/2}+\frac{(\alpha^{*2}-\beta^*)}{8\pi}\left(H^2+\frac{k}{a^2}\right)-\frac{\alpha^{*2}}{8\pi}\left(H^2+\frac{k}{a^2}\right)+\mathcal{O}(\alpha^{*3})\right) \nonumber \\
&+& \frac{2\pi\alpha^*\beta^*}{(\alpha^{*2}-\beta^*)^{5/2}}\left[\frac{(\alpha^{*2}-\beta^*)}{(4\pi)^{1/2}(\alpha^*+\sqrt{\alpha^{*2}-\beta^*})}\left(H^2+\frac{k}{a^2}\right)^{1/2}\right. \nonumber \\
&+&{\frac{\sqrt{\alpha^{*2}-\beta^*}}{\alpha^*+\sqrt{\alpha^{*2}-\beta^*}}\left(\frac{\alpha^*}{(4\pi)^{1/2}}\left(H^2+\frac{k}{a^2}\right)^{1/2}+\frac{(\alpha^{*2}-\beta^*)}{8\pi}\left(H^2+\frac{k}{a^2}\right)-\frac{\alpha^{*2}}{8\pi}\left(H^2+\frac{k}{a^2}\right)\right)}~ \nonumber \\
&-&\left.\frac{2(\alpha^{*2}-\beta^*)^{3/2}\alpha^*+2\alpha^{*4}-3\alpha^{*2}\beta^{*}+\beta^{*2}}{8\pi(2\alpha^{*2}+2\alpha^*\sqrt{\alpha^{*2}-\beta^*}-\beta^*)}\left(H^2+\frac{k}{a^2}\right)+\mathcal{O}(\alpha^{*3})\right] \nonumber \\
&=&\bcancel{\frac{2\pi (\alpha^{*2}+2\beta^*)}{3(\alpha^{*2}-\beta^*)^2}}-\bcancel{\frac{2\pi (\alpha^{*2}+2\beta^*)}{3(\alpha^{*2}-\beta^*)^2}} \nonumber \\
&+&\pi^{1/2}\left(\frac{\bcancel{\left(\alpha^{*3}-\alpha^*\beta^*\right)\left(\alpha^*+\sqrt{\alpha^{*2}-\beta^*}\right)}-\bcancel{\left(\alpha^{*3}-\alpha^*\beta^*\right)\left(\alpha^*+\sqrt{\alpha^{*2}-\beta^*}\right)}}{3\left(\alpha^{*2}-\beta^*\right)^2\left(\alpha^*+\sqrt{\alpha^{*2}-\beta^2}\right)}\right)\left(H^2+\frac{k}{a^2}\right)^{1/2} \nonumber \\
&+&\left(\frac{5}{6}+\frac{\bcancel{\left(\alpha^{*2}-\beta^*\right)^{5/2}\left(2\alpha^{*2}+2\alpha^*\sqrt{\alpha^{*2}-\beta^*}-\beta^*\right)}}{6\bcancel{\left(\alpha^{*2}-\beta^*\right)^{5/2}\left(2\alpha^{*2}+2\alpha^*\sqrt{\alpha^{*2}-\beta^*}-\beta^*\right)}}\right)\left(H^2+\frac{k}{a^2}\right) + \mathcal{O}(\alpha^*,\beta^*)
\nonumber \\
&=&H^2+\frac{k}{a^2}~.
\end{eqnarray}
In the last line in the above, after some straightforward but tedious algebra, all other terms cancel. To obtain first order GUP corrections in $\alpha^*$ and $\beta^*$ to the Friedmann equations, one must use the Taylor expansion up to fourth order to gather all required terms.

\end{appendices}
\newpage
%
%
%
%


\begin{thebibliography}{99}
%
%
%
%
%
\bibitem{GF}
A.~Zeilinger,
Rev. Mod. Phys. \textbf{71}, S288-S297 (1999).


\bibitem{CMW}
C.~M.~Will,
Living Rev. Rel. \textbf{17}, 4 (2014)
[arXiv:1403.7377 [gr-qc]].


\bibitem{Das:2008kaa}
S.~Das and E.~C.~Vagenas,
Phys. Rev. Lett. \textbf{101}, 221301 (2008)
[arXiv:0810.5333 [hep-th]].



\bibitem{Das:2010sj}
S.~Das and E.~C.~Vagenas,
Phys. Rev. Lett. \textbf{104}, 119002 (2010)
[arXiv:1003.3208 [hep-th]].



\bibitem{CM1}
J.~Magueijo,
Phys. Rev. D \textbf{73}, 124020 (2006)
[arXiv:gr-qc/0603073 [gr-qc]].



\bibitem{CM2}
A.~Hamma and F.~Markopoulou,
New J. Phys. \textbf{13}, 095006 (2011)
[arXiv:1011.5754 [gr-qc]].



\bibitem{CM3}
M.~M.~Dos Santos, T.~Oniga, A.~S.~Mcleman, M.~Caldwell and C.~H.~T.~Wang,
J. Plasma Phys. \textbf{79}, 437 (2013)
[arXiv:1301.0494 [quant-ph]].


\bibitem{CM4}
A.~Feller and E.~R.~Livine,
Class. Quant. Grav. \textbf{33}, no.6, 065005 (2016)
[arXiv:1509.05297 [gr-qc]].



\bibitem{CM5}
I.~Danshita, M.~Hanada and M.~Tezuka,
PTEP \textbf{2017}, no.8, 083I01 (2017)
[arXiv:1606.02454 [cond-mat.quant-gas]].



\bibitem{Buoninfante:2019fwr}
L.~Buoninfante, G.~G.~Luciano and L.~Petruzziello,
Eur. Phys. J. C \textbf{79}, no.8, 663 (2019)
[arXiv:1903.01382 [gr-qc]].


\bibitem{Blasone:2019wad}
M.~Blasone, G.~Lambiase, G.~G.~Luciano, L.~Petruzziello and F.~Scardigli,
Int. J. Mod. Phys. D \textbf{29}, no.02, 2050011 (2020)
[arXiv:1912.00241 [hep-th]].


\bibitem{CM6}
T.~W.~van de Kamp, R.~J.~Marshman, S.~Bose and A.~Mazumdar,
Phys. Rev. A \textbf{102}, no.6, 062807 (2020)
[arXiv:2006.06931 [quant-ph]].



\bibitem{CM7}
S.~A.~Haine,
New J. Phys. \textbf{23}, no.3, 033020 (2021)
[arXiv:1810.10202 [quant-ph]].




\bibitem{GUP1}
M.~Maggiore,
Phys. Lett. B \textbf{304}, 65-69 (1993)
[arXiv:hep-th/9301067 [hep-th]].



\bibitem{GUP2}
M.~Maggiore,
Phys. Lett. B \textbf{319}, 83-86 (1993)
[arXiv:hep-th/9309034 [hep-th]].



\bibitem{GUP3}
M.~Maggiore,
Phys. Rev. D \textbf{49}, 5182-5187 (1994)
[arXiv:hep-th/9305163 [hep-th]].




\bibitem{GUP4}
F.~Scardigli,
Phys. Lett. B \textbf{452}, 39-44 (1999)
[arXiv:hep-th/9904025 [hep-th]].



\bibitem{KMM}
A.~Kempf, G.~Mangano and R.~B.~Mann,
Phys. Rev. D \textbf{52}, 1108-1118 (1995)
[arXiv:hep-th/9412167 [hep-th]].



\bibitem{FC}
F.~Scardigli and R.~Casadio,
Class. Quant. Grav. \textbf{20}, 3915-3926 (2003)
[arXiv:hep-th/0307174 [hep-th]].



\bibitem{Das:2009hs}
S.~Das and E.~C.~Vagenas,
Can. J. Phys. \textbf{87}, 233-240 (2009)
[arXiv:0901.1768 [hep-th]].





\bibitem{Ali:2010yn}
A.~F.~Ali, S.~Das and E.~C.~Vagenas,
``The Generalized Uncertainty Principle and Quantum Gravity Phenomenology'',
[arXiv:1001.2642 [hep-th]].


\bibitem{Basilakos:2010vs}
S.~Basilakos, S.~Das and E.~C.~Vagenas,
JCAP \textbf{09}, 027 (2010)
[arXiv:1009.0365 [hep-th]].



\bibitem{ADV0}
A.~F.~Ali, S.~Das and E.~C.~Vagenas,
Phys. Rev. D \textbf{84}, 044013 (2011)
[arXiv:1107.3164 [hep-th]].


\bibitem{IPK}
I.~Pikovski, M.~R.~Vanner, M.~Aspelmeyer, M.~S.~Kim and C.~Brukner,
Nature Phys. \textbf{8}, 393-397 (2012)
[arXiv:1111.1979 [quant-ph]].


\bibitem{Scardigli:2014qka}
F.~Scardigli and R.~Casadio,
Eur. Phys. J. C \textbf{75}, no.9, 425 (2015)
[arXiv:1407.0113 [hep-th]].


\bibitem{SLV}
F.~Scardigli, G.~Lambiase and E.~Vagenas,
Phys. Lett. B \textbf{767}, 242-246 (2017)
[arXiv:1611.01469 [hep-th]].



\bibitem{KP}
S.~P.~Kumar and M.~B.~Plenio,
Phys. Rev. A \textbf{97}, no.6, 063855 (2018)
[arXiv:1708.05659 [quant-ph]].


\bibitem{VenezGrossMende}
   D.~Amati, M.~Ciafaloni and G.~Veneziano,
Phys. Lett. B \textbf{197}, 81 (1987); 
  D.~J.~Gross and P.~F.~Mende,
Phys. Lett. B \textbf{197}, 129-134 (1987); 
   D.~Amati, M.~Ciafaloni and G.~Veneziano,
Phys. Lett. B \textbf{216}, 41-47 (1989); 
    K.~Konishi, G.~Paffuti and P.~Provero,
Phys. Lett. B \textbf{234}, 276-284 (1990); 
    S.~Capozziello, G.~Lambiase and G.~Scarpetta,
Int. J. Theor. Phys. \textbf{39}, 15-22 (2000)
[arXiv:gr-qc/9910017 [gr-qc]].


\bibitem{Das:2020ujn}
A.~Das, S.~Das and E.~C.~Vagenas,
Phys. Lett. B \textbf{809}, 135772 (2020)
[arXiv:2006.05781 [gr-qc]].



\bibitem{CM8GS} 
G.~Lambiase and F.~Scardigli,
Phys. Rev. D \textbf{97}, no.7, 075003 (2018)
[arXiv:1709.00637 [hep-th]].

\bibitem{Iorio:2019wtn}
A.~Iorio, G.~Lambiase, P.~Pais and F.~Scardigli,
Phys. Rev. D \textbf{101}, no.10, 105002 (2020)
[arXiv:1910.09019 [hep-th]].

\bibitem{luciano}
L.~Petruzziello,
Class. Quant. Grav. \textbf{38}, no.13, 135005 (2021)
[arXiv:2010.05896 [hep-th]]; 
G.~G.~Luciano and L.~Petruzziello,
Eur. Phys. J. Plus \textbf{136}, no.2, 179 (2021).


\bibitem{CDS}
L.~Canetti, M.~Drewes and M.~Shaposhnikov,
New J. Phys. \textbf{14}, 095012 (2012)
[arXiv:1204.4186 [hep-ph]].



\bibitem{Sakh}
A.~D.~Sakharov,
Pisma Zh. Eksp. Teor. Fiz. \textbf{5}, 32-35 (1967).
 
 
 
\bibitem{27GL} 
J.~Dunkley \textit{et al.} [WMAP],
Astrophys. J. Suppl. \textbf{180}, 306-329 (2009)
[arXiv:0803.0586 [astro-ph]].



\bibitem{28GL} 
W.~M.~Yao \textit{et al.} [Particle Data Group],
J. Phys. G \textbf{33}, 1-1232 (2006).
 
 
 
\bibitem{CKB}
A.~G.~Cohen and D.~B.~Kaplan,
Nucl. Phys. B \textbf{308}, 913-928 (1988).



\bibitem{APS}
S.~H.~S.~Alexander, M.~E.~Peskin and M.~M.~Sheikh-Jabbari,
Phys. Rev. Lett. \textbf{96}, 081301 (2006)
[arXiv:hep-th/0403069 [hep-th]].
 
 
\bibitem{Cai:2008ys}
R.~G.~Cai, L.~M.~Cao and Y.~P.~Hu,
JHEP \textbf{08}, 090 (2008)
[arXiv:0807.1232 [hep-th]].

\bibitem{Zhu:2008cg}
T.~Zhu, J.~R.~Ren and M.~F.~Li,
Phys. Lett. B \textbf{674}, 204-209 (2009)
doi:10.1016/j.physletb.2009.03.020
[arXiv:0811.0212 [hep-th]].

\bibitem{AA}
A.~Awad and A.~F.~Ali,
JHEP \textbf{06}, 093 (2014)
[arXiv:1404.7825 [gr-qc]].

\bibitem{Giardino:2020myz}
S.~Giardino and V.~Salzano,
Eur. Phys. J. C \textbf{81}, no.2, 110 (2021)
[arXiv:2006.01580 [gr-qc]].
 
 
\bibitem{GTH}
G.~'t Hooft,
Conf. Proc. C \textbf{930308}, 284-296 (1993)
[arXiv:gr-qc/9310026 [gr-qc]].




\bibitem{LS}
L.~Susskind,
J. Math. Phys. \textbf{36}, 6377-6396 (1995)
[arXiv:hep-th/9409089 [hep-th]].




\bibitem{GUPAM}
P.~Bosso and S.~Das,
Annals Phys. \textbf{383}, 416-438 (2017)
[arXiv:1607.01083 [gr-qc]].




\bibitem{MV}
A.~J.~M.~Medved and E.~C.~Vagenas,
Phys. Rev. D \textbf{70}, 124021 (2004)
[arXiv:hep-th/0411022 [hep-th]].



\bibitem{ACAP}
G.~Amelino-Camelia, M.~Arzano and A.~Procaccini,
Phys. Rev. D \textbf{70}, 107501 (2004)
[arXiv:gr-qc/0405084 [gr-qc]].




\bibitem{LJG}
L.~J.~Garay,
Int. J. Mod. Phys. A \textbf{10}, 145-166 (1995)
[arXiv:gr-qc/9403008 [gr-qc]].




\bibitem{Adler:2001vs}
R.~J.~Adler, P.~Chen and D.~I.~Santiago,
Gen. Rel. Grav. \textbf{33}, 2101-2108 (2001)
[arXiv:gr-qc/0106080 [gr-qc]].





\bibitem{DC}
D.~Christodoulou,
Phys. Rev. Lett. \textbf{25}, 1596-1597 (1970).



\bibitem{CR}
D.~Christodoulou and R.~Ruffini,
Phys. Rev. D \textbf{4}, 3552-3555 (1971).



\bibitem{JB}
J.~D.~Bekenstein,
Phys. Rev. D \textbf{7}, 2333-2346 (1973).


 
 \bibitem{SH}
S.~W.~Hawking,
Phys. Rev. D \textbf{13}, 191-197 (1976).







\bibitem{KU1}
T.~Kugo and S.~Uehara,
Nucl. Phys. B \textbf{222}, 125-138 (1983).



\bibitem{KU2}
T.~Kugo and S.~Uehara,
Prog. Theor. Phys. \textbf{73}, 235 (1985).



\bibitem{gravbar} 
H.~Davoudiasl, R.~Kitano, G.~D.~Kribs, H.~Murayama and P.~J.~Steinhardt,
Phys. Rev. Lett. \textbf{93}, 201301 (2004)
[arXiv:hep-ph/0403019 [hep-ph]].





\bibitem{allGB}
N.~Azhar, A.~Jawad and S.~Rani,
Phys. Dark Univ. \textbf{32}, 100815 (2021);
H.~Davoudiasl,
Phys. Rev. D \textbf{88}, 095004 (2013)
[arXiv:1308.3473 [hep-ph]];
S.~D.~Odintsov and V.~K.~Oikonomou,
EPL \textbf{116}, no.4, 49001 (2016)
[arXiv:1610.02533 [gr-qc]];
S.~D.~Odintsov and V.~K.~Oikonomou,
Phys. Lett. B \textbf{760}, 259-262 (2016)
[arXiv:1607.00545 [gr-qc]];
V.~K.~Oikonomou and E.~N.~Saridakis,
Phys. Rev. D \textbf{94}, no.12, 124005 (2016)
[arXiv:1607.08561 [gr-qc]];
S.~Bhattacharjee and P.~K.~Sahoo,
Eur. Phys. J. C \textbf{80}, no.3, 289 (2020)
[arXiv:2002.11483 [physics.gen-ph]];
E.~H.~Baffou, M.~J.~S.~Houndjo, D.~A.~Kanfon and I.~G.~Salako,
Eur. Phys. J. C \textbf{79}, no.2, 112 (2019)
[arXiv:1808.01917 [gr-qc]];
M.~P.~L.~P.~Ramos and J.~P\'aramos,
Phys. Rev. D \textbf{96}, no.10, 104024 (2017)
[arXiv:1709.04442 [gr-qc]];
M.~Fukushima, S.~Mizuno and K.~i.~Maeda,
Phys. Rev. D \textbf{93}, no.10, 103513 (2016)
[arXiv:1603.02403 [hep-ph]];
H.~M.~Sadjadi,
Phys. Rev. D \textbf{76}, 123507 (2007)
[arXiv:0709.0697 [gr-qc]].



\bibitem{KT}
E.~W.~Kolb and M.~S.~Turner,
``The Early Universe,''
Front. Phys. \textbf{69}, 1-547 (1990).

\bibitem{Kinney:2006qm}
W.~H.~Kinney, E.~W.~Kolb, A.~Melchiorri and A.~Riotto,
Phys. Rev. D \textbf{74}, 023502 (2006)
[arXiv:astro-ph/0605338 [astro-ph]].


\bibitem{baryogenesis}
J.~M.~Cline,
``Baryogenesis,''
[arXiv:hep-ph/0609145 [hep-ph]];
A.~D.~Dolgov,
``CP violation in cosmology,''
[arXiv:hep-ph/0511213 [hep-ph]];
A.~Riotto,
``Theories of baryogenesis,''
[arXiv:hep-ph/9807454 [hep-ph]];
A.~Riotto and M.~Trodden,
Ann. Rev. Nucl. Part. Sci. \textbf{49}, 35-75 (1999)
[arXiv:hep-ph/9901362 [hep-ph]];
M.~Yoshimura,
J. Korean Phys. Soc. \textbf{29}, S236 (1996)
[arXiv:hep-ph/9605246 [hep-ph]];
G.~Lambiase, S.~Mohanty and A.~R.~Prasanna,
Int. J. Mod. Phys. D \textbf{22}, 1330030 (2013)
[arXiv:1310.8459 [hep-ph]].



\bibitem{Nenmeli:2021orl}
V.~Nenmeli, S.~Shankaranarayanan, V.~Todorinov and S.~Das,
Phys. Lett. B \textbf{821}, 136621 (2021)
[arXiv:2106.04141 [gr-qc]].



\bibitem{JKS} 
P.~Jizba, H.~Kleinert and F.~Scardigli,
Phys. Rev. D \textbf{81}, 084030 (2010)
[arXiv:0912.2253 [hep-th]].



\bibitem{Vagenas:2017fwa}
E.~C.~Vagenas, L.~Alasfar, S.~M.~Alsaleh and A.~F.~Ali,
Nucl. Phys. B \textbf{931}, 72-78 (2018)
doi:10.1016/j.nuclphysb.2018.04.004
[arXiv:1706.06502 [hep-th]].




\bibitem{vilasi} 
T.~Kanazawa, G.~Lambiase, G.~Vilasi and A.~Yoshioka,
Eur. Phys. J. C \textbf{79}, no.2, 95 (2019).



\bibitem{Aghababaei:2021gxe}
S.~Aghababaei, H.~Moradpour and E.~C.~Vagenas,
Eur. Phys. J. Plus \textbf{136}, 997 (2021)
doi:10.1140/epjp/s13360-021-02007-5
[arXiv:2109.14826 [gr-qc]].




\bibitem{Ali:2009zq}
A.~F.~Ali, S.~Das and E.~C.~Vagenas,
Phys. Lett. B \textbf{678}, 497-499 (2009)
[arXiv:0906.5396 [hep-th]].



\bibitem{Das:2010zf}
S.~Das, E.~C.~Vagenas and A.~F.~Ali,
Phys. Lett. B \textbf{690}, 407-412 (2010)
[erratum: Phys. Lett. B \textbf{692}, 342-342 (2010)]
[arXiv:1005.3368 [hep-th]].

\bibitem{Deb:2016psq}
S.~Deb, S.~Das and E.~C.~Vagenas,
Phys. Lett. B \textbf{755}, 17-23 (2016)
[arXiv:1601.07893 [gr-qc]].



\bibitem{CK}
R.~G.~Cai and S.~P.~Kim,
JHEP \textbf{02}, 050 (2005)
[arXiv:hep-th/0501055 [hep-th]].


\bibitem{Cai:2008gw}
R.~G.~Cai, L.~M.~Cao and Y.~P.~Hu,
Class. Quant. Grav. \textbf{26}, 155018 (2009)
[arXiv:0809.1554 [hep-th]].



\bibitem{SAH}
S.~A.~Hayward,
Class. Quant. Grav. \textbf{15}, 3147-3162 (1998)
[arXiv:gr-qc/9710089 [gr-qc]].


\end{thebibliography}
\end{document}